\documentclass[12pt]{article}

\usepackage{ifpdf}
\ifpdf
\usepackage{graphicx,color}
\usepackage{hyperref}
\else
\fi
\usepackage{amssymb,amsfonts,amsmath,cancel,cite,multirow}
\usepackage[capitalise]{cleveref}

\newcommand{\Tr}[0]{\mathrm{Tr}}
\newcommand{\tr}[0]{\mathrm{tr}}
\newcommand{\ave}[1]{\left \langle #1 \right \rangle}
\newcommand{\diag}{\text{diag}}

\newcommand{\Ocal}[0]{\mathcal{O}}
\newcommand{\Lcal}[0]{\mathcal{L}}
\newcommand{\MPl}[0]{M_{\text{Pl}}}

\newcommand{\lam}{\lambda}
\newcommand{\Lam}{\Lambda}
\newcommand{\ep}{\epsilon}
\newcommand{\hc}{\mathrm{h.c.}}
\newcommand{\del}{\delta}
\newcommand{\Del}{\Delta}
\newcommand{\si}{\sigma}
\newcommand{\Si}{\Sigma}
\newcommand{\ovl}{\overline}
\newcommand{\eqs}[1]{\begin{equation}\begin{split} #1 \end{split}\end{equation}}

\begin{document}

\begin{titlepage}

\begin{flushright}
	IPMU18-0195 \\
	CTPU-PTC-18-41
\end{flushright}

\vskip 1.35cm
\begin{center}

{\large
\textbf{
Ultraviolet Completion of a Composite Asymmetric Dark Matter Model with a Dark Photon Portal
}}
\vskip 1.2cm

Masahiro Ibe$^{a,b}$,
Ayuki Kamada$^c$,
Shin Kobayashi$^{a,b}$, \\
Takumi Kuwahara$^c$,
and Wakutaka Nakano$^{a,b}$

\vskip 0.4cm

\textit{$^a$
Kavli IPMU (WPI), UTIAS, The University of Tokyo, Kashiwa, Chiba 277-8583, Japan
}\\
\textit{$^b$
ICRR, The University of Tokyo, Kashiwa, Chiba 277-8582, Japan
} \\
\textit{$^c$
Center for Theoretical Physics of the Universe,
Institute for Basic Science (IBS), Daejeon 34126, Korea}

\vskip 1.5cm

\begin{abstract}

Composite asymmetric dark matter scenarios naturally explain why the dark matter mass density is comparable with the visible matter mass density.
Such scenarios generically require some entropy transfer mechanism below the composite scale; otherwise, their late-time cosmology is incompatible with observations.
A tiny kinetic mixing between a dark photon and the visible photon is a promising example of the low-energy portal.
In this paper, we demonstrate that grand unifications in the dark and the visible sectors explain the origin of the tiny kinetic mixing.
We particularly consider an ultraviolet completion of a simple composite asymmetric dark matter model, where asymmetric dark matter carries a $B-L$ charge.
In this setup, the longevity of asymmetric dark matter is explained by the $B-L$ symmetry,
while the dark matter asymmetry originates from the $B-L$ asymmetry generated by thermal leptogenesis.
In our minimal setup, the Standard Model sector and the dark sector are unified into $SU(5)_\mathrm{GUT} \times SU(4)_\mathrm{DGUT}$ gauge theories, respectively.
This model generates required $B-L$ portal operators while suppressing unwanted higher-dimensional operators that could wash out the generated $B-L$ asymmetry.
\end{abstract}

\end{center}
\end{titlepage}

\section{Introduction}

While astrophysical and cosmological observations have firmly established the existence of dark matter (DM), only a few properties of DM particles have been revealed: they should be stable or have a lifetime longer than the age of the Universe; and they interact with the standard model (SM) particles only weakly.

Stability is one of the critical ingredients to identify the nature of DM particles.
It is well known that the proton has a long lifetime, and its longevity is ensured by an accidental symmetry, called the baryon number symmetry, in the SM.
It is natural to consider that DM particles are stable for a similar reason: they are dark baryons in a strong dynamics in the dark sector, and the dark baryon number is conserved accidentally~\cite{Gudnason:2006yj, Dietrich:2006cm, Khlopov:2007ic, Khlopov:2008ty, Foadi:2008qv, Mardon:2009gw, Kribs:2009fy, Barbieri:2010mn, Blennow:2010qp, Lewis:2011zb, Appelquist:2013ms, Hietanen:2013fya, Cline:2013zca, Appelquist:2014jch, Hietanen:2014xca, Krnjaic:2014xza, Detmold:2014qqa, Detmold:2014kba, Asano:2014wra, Brod:2014loa, Antipin:2014qva, Appelquist:2015yfa, Appelquist:2015zfa, Antipin:2015xia, Co:2016akw, Dienes:2016vei, Ishida:2016fbp, Lonsdale:2017mzg, Berryman:2017twh, Mitridate:2017oky, Ibe:2018juk, Francis:2018xjd, Hardy:2014mqa, Hardy:2015boa, Bai:2018dxf,Braaten:2018xuw, Gresham:2018anj, Gresham:2017cvl, Gresham:2017zqi} (see Ref.~\cite{Kribs:2016cew} for a review).

Composite DM is particularly well motivated in the asymmetric dark matter (ADM) framework,
where the asymmetry generated in the visible and/or the dark sectors is communicated to the other sector
via some portal interactions (see Refs.~\cite{Nussinov:1985xr, Barr:1990ca, Barr:1991qn, Kaplan:1991ah, Dodelson:1991iv, Kuzmin:1996he, Fujii:2002aj, Kitano:2004sv, Farrar:2005zd, Gudnason:2006ug, Kitano:2008tk, Kaplan:2009ag} for early works and also Refs.~\cite{Davoudiasl:2012uw, Petraki:2013wwa, Zurek:2013wia} for reviews).
The common origin of the DM and the baryon asymmetries explains why the DM density today is about five times larger than the visible matter density when the DM mass is in the GeV range.
Dimensional transmutation of the strong dynamics can in turn naturally explain the DM mass at the GeV scale.
Besides, the dark baryons annihilate into dark pions quite efficiently, so that the asymmetric component dominates the DM relic density.
The dark baryon self-interaction mediated by dark mesons may also realize the velocity dependent cross section that addresses the dwarf galaxy-scale issues of structure formation of collisionless cold dark matter while leaves its success at galaxy clusters  (see Ref.~\cite{Tulin:2017ara} for a review).

In composite ADM scenarios, the dark sector is in thermal equilibrium with the visible sector through  high-energy portal interactions that communicate the asymmetry.
Accordingly, the dark sector possesses a sizable entropy density comparable to the one in the visible sector.
As the entropy densities are conserved separately in the two sectors after the decoupling of the high-energy portal interaction, the resultant dark sector entropy density is
carried over by the light particles in the dark sector such as dark pions.
The late-time energy density of the light particles could overclose the Universe or give a significant contribution to the dark radiation, depending on their masses~\cite{Blennow:2012de}.

To overcome such a shortcoming of composite ADM, one needs to introduce an additional low-energy portal that transfers the dark sector entropy density into the visible sector.
One promising candidate is a dark photon portal, where the dark photon decays into a pair of the electron and the positron through the kinetic mixing with the visible photon~\cite{Holdom:1985ag} (see also Refs.\cite{Alves:2009nf, Alves:2010dd, Ibe:2018juk} in the context of composite ADM).
It is shown that the MeV-scale dark photon with the kinetic mixing of the order of $10^{-9}$ is a viable portal evading all the experimental and cosmological constraints~\cite{Ibe:2018juk}.

On the other hand, the origin of the tiny kinetic mixing is unclear.
In this paper, we propose grand unifications (GUTs) in two sectors as its origin.
If the SM $U(1)_Y$ and the dark $U(1)_D$ dynamics are unified into separate non-abelian gauge dynamics at some high energies,
the kinetic mixing vanishes above the unification scales.
It arises through a higher dimensional non-renormalizable operator and is suppressed by the ratios between the GUT scales and the Planck scale.
For example, the dark GUT scale of the order of $10^{10}\,\mathrm{GeV}$ provides a kinetic mixing of the order of $10^{-9}$ when the unified gauge groups are broken by the vacuum expectation values (VEVs) of adjoint scalars.

This dark GUT scale is comparable with the minimal right-handed neutrino mass required for thermal leptogenesis, $M_{R} > 10^{9}\,\mathrm{GeV}$~\cite{Fukugita:1986hr} (see Refs.~\cite{Giudice:2003jh, Buchmuller:2005eh, Davidson:2008bu} for reviews).
Thus one can take advantage of right-handed neutrinos as the origins of the $B-L$ asymmetry and the high-energy portal operator.
As a particular example, we take a minimal composite ADM model proposed in Ref.~\cite{Ibe:2018juk}.
The model is based on dark quantum chromodynamics (dark QCD) and dark quantum electrodynamics (dark QED).
An ADM candidate in the model is dark nucleons that consist of dark quarks carrying $B-L$ number.
In this paper, we provide an ultraviolet (UV) completion of the simple ADM model proposed in Ref.~\cite{Ibe:2018juk} based on a product GUT of $SU(5)_\mathrm{GUT} \times SU(4)_\mathrm{DGUT}$.

Furthermore, the UV completion clarifies the origin of the high-energy portal operators with which the visible and the dark sectors share the $B-L$ asymmetries.
The portal interactions in composite ADM models are generically provided as non-renormalizable operators.
From the view point of the effective field theory, there is no reason why some of unwanted non-renormalizable operators which could wash out the $B-L$ asymmetry
are much suppressed than the required portal operators.
This issue can be addressed only by specifying the UV completions of the model of ADM.
Thus, it is important to construct viable UV completions of the composite ADM models in order to guarantee that such dangerous operators are safely neglected.

This paper is organized as follows.
In \cref{sec:review}, we briefly review a composite model of ADM~\cite{Ibe:2018juk}.
We construct UV models of a composite ADM model: a non-supersymmetric (non-SUSY) model in \cref{sec:nonsusy} and a SUSY model in \cref{sec:susy}.
\cref{sec:conclusion} is devoted to conclusions of our work.

%%%%%%%%%%%%%%%%%%%%%%%%%%%%%%%%%%%%%%%%%%%%%%%%%%%%%%%%%%%%%%%%%%%%%%%%%%%%%%%%%
\section{Simple Composite ADM Model \label{sec:review}}
In this section, we sketch out a composite ADM model and fix our notation.
We consider the vector-like two-flavor $SU(3)_D \times U(1)_D$ dynamics proposed in Ref.~\cite{Ibe:2018juk}.
The dark quarks consist of (anti-)fundamental representations in $SU(3)_D$, which have the same $B-L$ numbers as the up-type and the down-type quarks in the visible QCD.
We list the minimal particle contents in the dark sector for the model in \cref{tab:Charge_dark}.
The chiral symmetry of the dark quarks are softly broken by the current quark masses,
\eqs{
\label{eq:mass}
\mathcal{L}_\mathrm{mass} = m_{U'} \ovl{U}' U' + m_{D'} \ovl{D}' D' + \hc
}
The dark quarks are confined into dark mesons and dark baryons below the dynamical scale of $SU(3)_D$, i.e., $\Lam_{\mathrm{DQCD}}$.
Dark baryons carry $B-L$ numbers, and thus the lightest one is stable in the dark sector and is a good candidate for DM.

The lightest dark mesons are also stable in composite models, and hence, they could lead to the overclosure of the Universe or a too large effective number of neutrino species,
$N_\mathrm{eff}$.
The dark QED is introduced in order to avoid the cosmological problems.
When the dark quarks are charged under $U(1)_D$, the dark mesons annihilate into the dark photons.
The $U(1)_D$ charges for the dark quarks are determined by the required existence of a neutral dark baryon, which is essential for the high-energy portal operator described below.

\begin{table}
	\centering
	\caption{Charge assignment of the dark quarks for a composite ADM model.
	$SU(3)_D$ and $U(1)_D$ are gauge symmetries of the dark sector, while $U(1)_{B-L}$ is the global symmetry
	shared with the visible sector.
	}
	\label{tab:Charge_dark}
	\begin{tabular}{|c|c|c||c|}
		\hline
		& $SU(3)_D$ & $U(1)_D$ & $U(1)_{B-L}$ \\ \hline
		$U'$ & $\mathbf{3}$ & $2/3$ & $1/3$ \\
		$\ovl U'$ & $\ovl{\mathbf{3}}$ & $-2/3$ & $- 1/3$ \\
		$D'$ & $\mathbf{3}$ & $-1/3$ & $1/3$ \\
		$\ovl D'$ & $\ovl{\mathbf{3}}$ & $1/3$ & $- 1/3$ \\ \hline
	\end{tabular}
\end{table}

The dark photon can decay into the visible particles when the kinetic mixing with the visible photon and a mass of the dark photon are introduced,
\eqs{
\Lcal_{\gamma^\prime} = \frac\ep2 F_{\mu\nu} F^{\prime\mu\nu} + \frac{m_{\gamma^\prime}^2}{2} A_\mu^\prime A^{\prime \mu} \,.
}
Here, $F_{\mu\nu}$ and $F^\prime_{\mu\nu}$ are the field strengths of the SM photon $A_\mu$ and the dark photon $A^\prime_\mu$, respectively.
The dark photon parameters $(\ep, m_{\gamma^\prime})$ are severely constrained by beam dump experiments, collider experiments~\cite{Bauer:2018onh}, SN 1987A~\cite{Chang:2016ntp, Chang:2018rso}, and the effects on the effective number of neutrino species $N_{\mathrm{eff}}$~\cite{Ibe:2018juk}.
In this work we assume the viable dark photon parameters in the ranges of $\ep = 10^{-10} \, \text{--} \, 10^{-9}$ and $m_{\gamma^\prime} = \mathcal{O}(10^2 \, \text{--}\, 10^3) \, \text{MeV}$~\cite{Ibe:2018juk}.
The origin of the tiny kinetic mixing is unclear, while is naturally understood in a GUT model investigated in the next section.

In the model proposed in Ref.~\cite{Ibe:2018juk}, the $B-L$ asymmetry in the visible sector is assumed to be generated by thermal leptogenesis.
The right-handed neutrinos couple to the SM lepton doublet $L$ and the SM Higgs doublet $H$ as
\eqs{
\Lcal_N = \frac{M_R}{2} \ovl N \ovl N + y_N L H \ovl N + \hc \,,
}
with Majorana masses being $M_R \gtrsim 10^{9}~\mathrm{GeV}$.
Remark that the right-handed neutrinos can also generate tiny neutrino masses through the see-saw mechanism~\cite{Minkowski:1977sc, Yanagida:1979as, GellMann:1980vs, Glashow:1979nm}.
It relates the Yukawa coupling with the observed neutrino mass $m_{\nu}$ as
\eqs{
y_{N}^{2} \sim 10^{-5} \left( \frac{m_{\nu}}{0.1 \, \mathrm{eV}} \right) \left( \frac{M_{R}}{10^{9} \, \mathrm{GeV}} \right) \,.
\label{eq:see-saw}
}
The generated $B-L$ asymmetry is shared between the dark and the visible sectors through portal operators,
\eqs{
\Lcal_{\text{portal}} =
\frac{c_1 y_N}{\Lam^2 M_R} (\ovl U' \ovl D' \ovl D') (L H)
+ \frac{c_2 y_N}{\Lam^2 M_R} (U'^\dag D'^\dag \ovl D') (L H) + \hc \,,
\label{eq:portal1_pheno}
}
which are obtained from
\eqs{
\Lcal_{\text{portal}} = \frac{c_1}{\Lam^2} (\ovl U' \ovl D' \ovl D') \ovl N
+ \frac{c_2}{\Lam^2} (U'^\dag D'^\dag \ovl D') \ovl N + \hc \,,
\label{eq:portal2_pheno}
}
below the energy scale of $M_R$.
Here $\Lam$ is the portal scale and $c_1$ and $c_2$ are constants.

After a part of $B-L$ is stored in the dark sector, the dark nucleons form below the dark confinement scale, and then the lightest one is the ADM candidate.
The lightest one is almost stable since the $B-L$ number is approximately conserved in the dark sector and the portal interactions to visible matters are suppressed by $\Lam$.
The dark neutron, which consists of $U'D'D'$ and $\ovl U' \ovl D' \ovl D'$, and the dark proton, which consists of $U'U'D'$ and $\ovl U' \ovl U' \ovl D'$ are ADM candidates in this setup.
The mass of ADM particles is determined by the ratio of the asymmetries in the SM and the dark sectors: $m_{\mathrm{DM}} = 8.5~\mathrm{GeV}$~\cite{Ibe:2011hq,Fukuda:2014xqa,Ibe:2018juk}.
It implies that the dark dynamical scale $\Lam_{\mathrm{DQCD}}$ is an order of magnitude larger than the QCD scale, namely $\Lam_{\mathrm{DQCD}} \sim 2~\mathrm{GeV}$.

The portal scale $\Lam$ is bounded from below by neutrino flux measurements~\cite{Fukuda:2014xqa}.
Meanwhile the portal interactions should decouple after the $B-L$ asymmetry is generated, namely, the decoupling temperature should be below $M_R$.
The decoupling temperature is estimated as $T_{*} \sim M_{*} (M_{*} / M_\mathrm{Pl})^{1/5}$ with $M_{*}$ collectively denoting $(\Lam^2 M_R / c_i y_N)^{1/3}$ ($i = 1, 2$) and $\MPl = 2.4 \times 10^{18}~\mathrm{GeV}$ being the reduced Planck mass.
Noting \cref{eq:see-saw} and requiring $M_R < \Lam$ for the consistency of the renormalizable operator, one obtains
\eqs{
M_R < \Lam \lesssim 10 ~ c_{i}^{1/2} M_R \,, ~~~
\text{and thus}~~~ c_{i} \gtrsim 0.01 \,.
\label{eq:pheno_constraint}
}
The origin of the portal operators can be easily explained if a scalar field charged under $SU(3)_D$ with a mass about $\Lam$ is introduced.
In the next section we will construct a $SU(4)_{\mathrm {DGUT}}$ unified model of $SU(3)_D$ and $U(1)_D$ gauge dynamics in the dark sector.

As the $B-L$ symmetry is softly broken by the right-handed neutrino masses, no symmetry prohibits the operators that carry different $B-L$ charges so far.
If the following operators are also relevant after the $B-L$ asymmetry is generated, the asymmetry is washed out:
\eqs{
\Lcal = \frac{c_1^\prime}{\Lam^{\prime 2}} (U' D' D') \ovl N
+ \frac{c_2^\prime}{\Lam^{\prime 2}} (\ovl U'^\dag \ovl D'^\dag D') \ovl N + \hc
}
We approximately obtain the cutoff scale $\Lam^\prime$ should be larger than $10^{11}~\mathrm{GeV}$ by requiring that the decoupling temperature of the interactions should be higher than $M_R \sim 10^{9}~\mathrm{GeV}$.
The hierarchy between $\Lam$ and $\Lam^\prime$, namely, between the desirable operators and dangerous operators for an ADM scenario, is also understood in a natural way by assuming a GUT model in the dark sector.%
\footnote{When $\bar{N}$'s are integrated out, the dark neutrons $n'\propto \ovl{U}'\ovl{D}'\ovl{D}'$ obtains a tiny $B-L$ breaking mass of ${\cal O}(\Lambda_{\mathrm{ DQCD}}^6/M_R \Lambda^5)$.
The $B-L$ breaking mass is, however, small enough not to washout the $B-L$ asymmetry. }

\section{Non-Supersymmetric Realization \label{sec:nonsusy}}

We consider a non-SUSY $SU(5)_{\mathrm{GUT} } \times SU(4)_{\mathrm{DGUT}}$ GUT model.
Here, $SU(5)_{\mathrm{GUT}}$ stands for the grand unified gauge group of the SM sector in the Georgi-Glashow model~\cite{Georgi:1974sy}.
The $SU(5)_{\mathrm{GUT}}$ gauge symmetry is spontaneously broken down to the SM gauge group $SU(3) \times SU(2) \times U(1)$ by the VEV of an adjoint scalar field $\Si(\mathbf{24})$, i.e., $\ave{\Si} = v_{24} ~ \diag(2,2,2,-3,-3)$.
We note that the GUT scale $v_{24}$ in the visible sector is assumed to be of the order of $10^{16}~\mathrm{GeV}$ to avoid too rapid proton decay~\cite{Miura:2016krn} even though intermediate-scale SUSY is not introduced.
Such a large GUT scale is achieved if some additional fields are introduced at an intermediate scale (see Refs.~\cite{Murayama:1991ah, Dorsner:2005fq, Bajc:2006ia, Bajc:2007zf,
Ibe:2009gt,Aizawa:2014iea,Cox:2016epl}).

We take the minimal option for the dark sector, $SU(4)_{\mathrm{DGUT}}$, which includes $SU(3)_D \times U(1)_D$ as a subgroup.
The $SU(4)_{\mathrm {DGUT}}$ symmetry is broken by an adjoint scalar field $\Xi'(\mathbf{15})$ by its VEV, i.e., $\ave{\Xi'} = v_{15} ~\diag(1,1,1,-3)$.
We assume that $v_{15}$ is of the order of $10^{10}~\mathrm{GeV}$ and is much smaller than $v_{24}$.
We note that the dark sector is an asymptotically free theory, and hence, the perturbativity in the dark sector is ensured up to the Planck scale.

\subsection{Tiny Kinetic Mixing}

The smallness of the visible photon-dark photon kinetic mixing $\ep$ is naturally explained in this setup.
We assume that any non-renormalizable operator is suppressed by the reduced Planck mass $\MPl$ above the $SU(5)_{\mathrm{GUT}} \times SU(4)_{\mathrm {DGUT}}$ GUT scale.
Under this assumption, the kinetic mixing arises from the following operator:
\eqs{
\Lcal_{\ep} = \frac{1}{\MPl^2} \tr(F_{G\,\mu\nu} \Si) \Tr(F^{\mu\nu}_D \Xi') \, ,
\label{eq:nonren-mixing}
}
where $\tr$ and $\Tr$ denote traces of $SU(5)_{\mathrm{GUT}}$ and $SU(4)_{\mathrm {DGUT}}$ indices, respectively.
$F_{G\,\mu\nu}$ and $F^{\mu\nu}_D$ are the field strengths of $SU(5)_{\mathrm {GUT}}$ and $SU(4)_{\mathrm{DGUT}}$, respectively.
Below the GUT scale, the kinetic mixing is given by%
\footnote{Radiative contributions also arise via three-loop diagrams involving right-handed neutrinos.
They are suppressed by loop factors and small $y_{N}$ [see \cref{eq:see-saw}], and thus are subdominant.}
\eqs{
\ep = \frac{2\sqrt{10} v_{24} v_{15} \cos \theta_W}{M_\mathrm{Pl}^2}
\simeq 10^{-9} \left( \frac{v_{24}}{2 \times 10^{16}~\mathrm{GeV}} \right) \left( \frac{v_{15}}{5 \times 10^{10}~\mathrm{GeV}} \right) \, ,
\label{eq:epsilon}
}
with the Weinberg angle being $\sin^2\theta_W \simeq 0.23$.
We naturally obtain  the tiny mixing parameter $\ep$ thanks to the hierarchy of $v_{15} \ll v_{24} < \MPl$.

\subsection{$B-L$ Portal Operator \label{sec:portal}}
In the GUT picture, $U(1)_{B-L}$ is realized as the ``fiveness'' $U(1)_5$ that commutes with the $SU(5)_{\mathrm {GUT}}$ and the $SU(4)_{\mathrm{DGUT}}$ symmetries.
The fiveness charge $Q_5$ is related to the hypercharge $Y$ and the $U(1)_D$ charge $D$  via $Q_5 = 5(B-L) - 4 Y - \frac52 D$.
\cref{tab:Q5Chargeassign} shows the minimal particle contents of $SU(4)_{\mathrm {DGUT}}$ and their charge assignment.

The dark quarks listed in \cref{tab:Charge_dark} are unified into the $\mathbf{6}$\,, $\mathbf{4}$\,,
and $\ovl{\mathbf{4}}$ representations of $SU(4)_{\mathrm{DGUT}}$ that are denoted by $Q_U\, , Q_D$\,, and $\ovl Q_{\ovl D}$\,, respectively.
Indeed, under the symmetry breaking of $SU(4)_{\mathrm{DGUT}} \to SU(3)_D \times U(1)_D$, a fundamental representation in $SU(4)_{\mathrm {DGUT}}$ is decomposed as $\mathbf{4} \to \mathbf{3}_{-1/3} + \mathbf{1}_{1}$, while an anti-symmetric representation in $SU(4)_{\mathrm{DGUT}}$ is decomposed as $\mathbf{6} \to \mathbf{3}_{2/3} +  \ovl{\mathbf{3}}_{-2/3}$:
\eqs{
Q'_U & = \frac{1}{\sqrt2} \left(
\begin{array}{cccc}
	0 & \ovl U'^3 & - \ovl U'^2 & U'_1 \\
	- \ovl U'^3 & 0 & \ovl U'^1 & U'_2 \\
	\ovl U'^2 & - \ovl U'^1 & 0 & U'_3 \\
	-U'_1 & - U'_3 & - U'_3 & 0 \\
\end{array}
\right) \,, ~~~~~
Q'_D = \left(
\begin{array}{c}
	D'_1 \\
	D'_2 \\
	D'_3 \\
	\ovl E' \\
\end{array}
\right) \, , ~~~~~
\ovl Q'_{\ovl D} = \left(
\begin{array}{c}
	\ovl D'^1 \\
	\ovl D'^2 \\
	\ovl D'^3 \\
	E' \\
\end{array}
\right) \, .
}
Here, $E'$ ($\ovl E'$) is a $SU(3)_D$ singlet with $U(1)_D$ charge $-1$ (+1), and thus we refer to them as the dark electron.
The sub- and superscripts denote $SU(3)_D$ indices.

We also introduce a fundamental scalar field $H'$, which is decomposed into a dark-colored Higgs triplet $\phi_C$ and a dark $U(1)_D$ breaking Higgs $\phi_D$.%
\footnote{It is possible to introduce another representation for the dark $U(1)_D$ breaking Higgs and the dark-colored Higgs triplet.
For instance, a symmetric representation is decomposed as $\mathbf{10} \to \mathbf{6}_{-2/3} + \mathbf{3}_{2/3} + \mathbf{1}_{2}$.
Therefore, the $B-L$ portal interactions do not arise from the dark colored Higgs triplet from the $\mathbf{10}$ representation without introducing extra fermions.
}
We impose fine-tuning of parameters in order to realize the mass difference between $\phi_C$ and $\phi_D$, namely between $10^{10}$~GeV and 1~GeV.

\begin{table}
	\centering
	\caption{
	Charge assignment of fermions and scalars in the minimal $SU(5)_{\mathrm {GUT}} \times SU(4)_{\mathrm {DGUT}}$ unified model.
	The upper rows of the tables show the assignment in $SU(5)_{\mathrm {GUT}}$ sector while the lower rows show those in $SU(4)_{\mathrm {DGUT}}$ sector.
	}
	\label{tab:Q5Chargeassign}
	\begin{tabular}{|c|c|c|c|}
		\hline
		& $SU(5)_{\mathrm {GUT}}$ & $SU(4)_{\mathrm {DGUT}}$ & $U(1)_5$ \\ \hline
		$\Psi_i$ & $\mathbf{10}$ & $\mathbf{1}$ & $1$ \\
		$\Phi_i$ & $\ovl{\mathbf{5}}$ & $\mathbf{1}$ & $-3$ \\
		$\ovl N_i$ & $\mathbf{1}$ & $\mathbf{1}$ & $5$ \\ \hline
		$Q'_U$ & $\mathbf{1}$ & $\mathbf{6}$ & $0$ \\
		$Q'_D$ & $\mathbf{1}$ & $\mathbf{4}$ & $5/2$ \\
		$\ovl Q'_{\ovl D}$ & $\mathbf{1}$ & $\ovl{\mathbf{4}}$ & $- 5/2$ \\ \hline
	\end{tabular}
	\begin{tabular}{|c|c|c|c|}
		\hline
		& $SU(5)_{\mathrm{GUT}}$ & $SU(4)_{\mathrm{DGUT}}$ & $U(1)_5$ \\ \hline
		$H$ & $\mathbf{5}$ & $\mathbf{1}$ & $2$ \\
		$\Si$ & $\mathbf{24}$ & $\mathbf{1}$ & $0$ \\ \hline
		$H'$ & $\mathbf{1}$ & $\mathbf{4}$ & $-5/2$ \\
		$\Xi'$ & $\mathbf{1}$ & $\mathbf{15}$ & $0$ \\ \hline
	\end{tabular}
\end{table}

Let us consider the generic Lagrangian density that is invariant under $SU(5)_{\mathrm{GUT}} \times SU(4)_{\mathrm {DGUT}}$.
We also assume that the global ``fiveness'' $U(1)_5$ is softly broken by Majorana masses $M_R$ for $\ovl N$.%
\footnote{$U(1)_5$ can be a gauge symmetry since the gauge anomalies are cancelled thanks to the right-handed neutrinos.
When we consider the gauged $U(1)_5$, $M_R$ is generated from a VEV of a $U(1)_5$ breaking scalar field.
}
Yukawa interactions for dark fermions are given by
\eqs{
\Lcal_{\text{Yukawa}} = - Y_D \ep^{\alpha\beta\gamma\del} H'_{\alpha} Q'_{U[\beta\gamma]} Q'_{D \del}
- Y_{\ovl D} H'^{\dag \alpha} Q'_{U[\alpha \beta]} \ovl Q'_{\ovl D}{}^\beta
- Y_N H'_{\alpha} \ovl Q_{\ovl D}'{}^\alpha \ovl N + \hc \, ,
}
where the Greek letters $\alpha,\beta,\cdots = 1, \cdots, 4$ are $SU(4)_{\mathrm {DGUT}}$ indices and $\ep^{\alpha\beta\gamma\del}$ is the totally antisymmetric tensor of $SU(4)_{\mathrm {DGUT}}$.
A square bracket $[\dots]$ represents antisymmetric indices.

Below the energy scale of the mass of $\phi_C$, denoted by $M_C$, the relevant effective Lagrangian density for portal interactions is given by
\eqs{
\Lcal_{\text{portal}} & = \frac{Y_N Y_{\ovl D}}{\sqrt2 M_C^2} \ep_{abc} (\ovl U'^a \ovl D'^b) (\ovl D'^c \ovl N)
- \frac{Y_N Y_D^\ast}{\sqrt2 M_C^2} \ep_{abc} (U'^{\dag a} D'^{\dag b}) (\ovl D'^c \ovl N)
+ \hc \,,
}
or below the energy scale of $M_R$,
\eqs{
\Lcal_{\text{portal}} & = - \frac{Y_N Y_{\ovl D} y_N}{\sqrt2 M_C^2 M_R} \ep_{abc} (\ovl U'^a \ovl D'^b) \ovl D'^c (LH)
+ \frac{Y_N Y_D^\ast y_N}{\sqrt2 M_C^2 M_R} \ep_{abc} (U'^{\dag a} D'^{\dag b}) \ovl D'^c (LH)
+ \hc
\label{eq:portal_su4}
}
Here, $a, b, c = 1, 2, 3$ represent $SU(3)_D$ indices and $\ep_{abc}$ is the totally antisymmetric tensor of $SU(3)_D$.
These are exactly what transfers the $B - L$ asymmetry generated by thermal leptogenesis into the dark sector [see \cref{eq:portal1_pheno,eq:portal2_pheno}].
Although the massive gauge bosons associated with $SU(4)_{\mathrm {DGUT}}$ breaking also give rise to four-Fermi operators, they are irrelevant for the portal interactions.
There arise no harmful portal operators that could wash out the $B - L$ asymmetry in combination with \cref{eq:portal_su4}.
The phenomenological constraint \cref{eq:pheno_constraint} reads
\eqs{
M_R < M_C \lesssim 10 ~ (Y_N Y_{D , \ovl D} )^{1/2} M_R \,,
~~~\text{and thus}~~~ Y_N Y_{D , \ovl D} \gtrsim 0.01 \,.
\label{eq:Y_constraint}
}
We remark that $M_C$, which is expected to be of order of $v_{15} = \Ocal(10^{10})~\mathrm{GeV}$, is larger than $M_{R} \sim 10^{9} ~\mathrm{GeV}$ required for thermal leptogenesis.

Dark nucleons can decay via the massive gauge bosons from $SU(4)_{\mathrm {DGUT}}$ when dark electron $E'$ is lighter than the dark nucleons.
Since the dark GUT scale is much lower than the visible GUT scale and the dark dynamical scale $\Lam_{\mathrm{DQCD}}$ is 10 times larger than the QCD scale, the DM decays within the age of the Universe:
\eqs{
\tau (n^\prime \to E' + \pi^\prime) \sim \frac{M_{\mathrm{DGUT}}^4}{g_U^{\prime 4} \Lam_{\mathrm{DQCD}}^5}
\sim 5 \times 10^7  ~\mathrm{yr} \left( \frac{M_{\mathrm{DGUT}}}{10^{10}~\mathrm{GeV}} \right)^4 \left( \frac{2~\mathrm{GeV}} {\Lam_{\mathrm{DQCD}}}\right)^5 \,,
\label{eq:nucdecay}
}
where $g_U^\prime$ is the $SU(4)_{\mathrm {DGUT}}$ gauge coupling at the scale of $M_{\mathrm{DGUT}}$.
The $SU(3)_D$ fine-structure constant $\alpha_S^{\prime -1}$ vanishes at the dark dynamical scale $\Lam_{\mathrm{DQCD}} \sim 2\,\mathrm{GeV}$,
which determines the value of $\alpha_S'$ at the dark GUT scale, i.e., $M_{\mathrm{DGUT}}$.
We obtain $g_U^{\prime 2} \sim 0.38$ assuming $M_{\mathrm{DGUT}} \sim 10^{10}~\mathrm{GeV}$ and $\Lam_{\mathrm{DQCD}} \sim 2~\mathrm{GeV}$.

To avoid this problem, we thus assume that the dark electron obtains a heavy mass via $SU(4)_{\mathrm {DGUT}}$ symmetry breaking for simplicity, that is, $m_E' \gg m_{p',n'}$.
In this case, the tiny dark quark current mass in \cref{eq:mass} requires fine-tuning between the vector-like mass term $Q_D' \ovl Q_{\ovl D}'$ and the Yukawa coupling $Q_D' \Xi' \ovl Q_{\ovl D}'$,
while giving the dark electron masses of the order of $v_{15}$.

\subsection{Dark Matter Phenomenology}
Last but not least, we discuss the decay of the heavier dark nucleon.
The heavier dark nucleon decays into the lighter one by emitting the dark photon in this model when the mass difference between the dark nucleons is larger than the mass of the dark photon, $m_{\gamma^\prime}$.
This process is induced by the mixing between the dark pion and the $U(1)_D$ breaking Higgs.
In this paper, we consider the case that this decay channel is open for simplicity.%
\footnote{When the mass difference is smaller than $m_{\gamma^\prime}$, the heavier dark nucleon decays into the lighter one with a pair of electron and positron via the dark photon-visible photon mixing instead decay with emitting the dark photon \cite{Ibe:2018juk}.}

Let us consider the interaction among the dark nucleons and the dark photon.
At the leading order the dark pions couple to the axial dark nucleon current, and the interaction is given by
\eqs{
\Lcal_{\pi^\prime N^\prime} =  - \frac{g_A}{f_{\pi^\prime}} D_\mu \pi^{\prime a} \left(N^{\prime\dag} \ovl \si^\mu \tau^a N^\prime - \ovl N^{\prime} \si^\mu \tau^a \ovl N^{\prime\dag}\right) \,,
\label{eq:piNint}
}
where $g_A$ and $f_{\pi^\prime}$ are an axial coupling constant of the dark nucleons and a dark pion decay constant, respectively.
$D_\mu$ is the covariant derivative of the dark pions.
$N^\prime$ and $\ovl N^\prime$ are dark isospin doublets, and the dark pion multiplet $\pi^\prime$ is defined as
\eqs{
\pi^{\prime a} \tau^a & = \frac1{\sqrt2} \left(
\begin{array}{cc}
	\pi^{\prime 0} & \sqrt2 \pi^{\prime +} \\
	\sqrt2 \pi^{\prime -} & - \pi^{\prime 0}
\end{array}
\right) \,,
}
where $\tau^a$ are the generators of dark isospin $SU(2)$.
The superscripts indicate the $U(1)_D$ charges.

When there are parity violating masses for the dark quarks, the $U(1)_D$ charged dark-pion $\pi^{\prime +}$ gets a VEV.
The chiral Lagrangian density for the dark pions is given by
\eqs{
\Lcal_{\chi\text{PT}} & = \frac{f_{\pi^\prime}^2}{4} \tr (\partial_\mu U_{\pi'} \partial^\mu  U_{\pi'}^\dag)
+ \left[ B \tr (M  U_{\pi'}) + \hc \right] + \cdots \,, \\
U_{\pi'} & = \exp \left( \frac{i \pi^{\prime a} \tau^a}{f_{\pi^\prime}} \right) \,, ~~~~~
M = \left(
\begin{array}{cc}
 m_{U'} & Y_D v_D \\
 Y_{\ovl D} v_D & m_{D'}
\end{array}
\right) \,,
}
with the dark quark current masses $m_{U'}$ and $m_{D'}$, and the $U(1)_D$ breaking Higgs VEV $v_D$.
$B$ is a dimensionful parameter of the order of $f_{\pi^\prime}^3$.
Expanding $U_{\pi'}$, we obtain the dark pion mass term,
\eqs{
\Lcal_{\pi^2} & = i \Del m_{\pi^\prime}^{\ast 2} v_D \pi^{\prime +}
- i \Del m_{\pi^\prime}^2 v_D \pi^{\prime -}
- \frac12 m_{\pi^\prime}^2 (\pi^{\prime 0})^2
- m_{\pi^\prime}^2 \pi^{\prime +}\pi^{\prime -} \,,
}
where
\eqs{
\Del m_{\pi^\prime}^2 \equiv
\frac{B}{f_{\pi^\prime}} (Y^\ast_{\ovl D} -Y_D) \, , ~~~~~
m_{\pi^\prime}^2 \equiv
\frac{B (m_{U'} + m_{D'})}{2 f_{\pi^\prime}^2} \, .
}
The dark pion gets a VEV,
\eqs{
\ave{\pi^{\prime +}} \equiv i v_\pi = \frac{i \Del m_{\pi^\prime}^2}{m_{\pi^\prime}^2} v_D \,.
}

Here, we implicitly assume that the $U(1)_D$ breaking masses $Y_{D , \ovl D} ~ v_D$ are subdominant when compared to the current masses.
Otherwise, the typical scale of the dark Higgs and the dark pions is expected to be of the order of the dark QCD scale since $\Del m_{\pi^\prime}^2 \sim \Ocal(f_{\pi^\prime}^2)$ dominates their masses.
In our scenario, the dark QCD scale is around 2~GeV, and thus the chiral symmetry is expected to be a good symmetry if the dark quark current masses are below 200\,MeV.
When we assume that the mass of the dark photon is of order of $100\,\text{MeV}$, $Y_{D , \ovl D} ~ v_D$ is approximately given by
\eqs{
Y_{D , \ovl D} ~ v_D \sim 11\,\text{MeV} \left(\frac{Y_{D , \ovl D}}{0.1} \right)
 \left( \frac{7 \times 10^{-2}}{\alpha^\prime} \right)^{1/2}
 \left(\frac{m_{\gamma^\prime}}{100\,\text{MeV}} \right) \,.
}
Thus, without assuming very tiny couplings $Y_{D , \ovl D}$,%
\footnote{Indeed, we cannot make $Y_{D , \ovl D}$ tiny so that the $B - L$ asymmetry is efficiently transferred after thermal leptogenesis [see \cref{eq:Y_constraint}].}
we can achieve the $U(1)_D$ breaking masses which are an order of magnitude smaller than the dark quark current masses.

Here, we use the $U(1)_D$ fine-structure constant $\alpha^\prime$ estimated in the GUT framework in the following way.
As $U(1)_D$ and $SU(3)_D$ are unified into $SU(4)_{\mathrm {DGUT}}$ at the dark GUT scale, $\alpha^\prime$ and $\alpha^\prime_S$ are identified there.
Therefore, the low-energy value of $\alpha^\prime$ is given by,
\eqs{
\alpha^{\prime -1}(\Lam_{\mathrm{DQCD}}) =
\frac{8}{3} \frac{(b^\prime - b_s^\prime)}{2 \pi} \ln \left(\frac{M_{\mathrm{DGUT}}}{\Lam_{\mathrm{DQCD}}} \right) \, .
\label{eq:U1Dcoupling}
}
Here, $b^\prime = 23/24$ and $b_s^\prime = -29/3$ are the one-loop $\beta$ function coefficients of $U(1)_D$ and $SU(3)_D$ gauge couplings, respectively.
The prefactor of $3/8$ arises from the $SU(4)_{\mathrm {DGUT}}$ normalization.

After the dark pion gets a VEV, the dark photon gets its mass not only via the dark Higgs but also via the dark pion.
The Nambu-Goldstone boson eaten by the dark photon is a mixture of the phase degrees of freedoms of them, and therefore the interaction in \cref{eq:piNint} leads to the interaction between the dark photon and the dark nucleons,
\eqs{
\Lcal_{\gamma^\prime p^\prime n^\prime} & =
- \frac{g_A g_D v_{\pi^\prime} }{f_{\pi^\prime}} A_\mu^\prime (- p^{\prime\dag} \ovl \si^\mu n^\prime + \ovl p^\prime \si^\mu \ovl n^{\prime \dag}) + \hc
}
This interaction leads to the prompt decay of the heavier dark nucleon to the lightest one in this model, when the mass difference between the dark
nucleons is larger than $m_{\gamma'}$.

The dark proton can interact with the SM proton via the kinetic mixing, and hence a constraint from the DM direct detection experiment is much stronger than other constraints unless DM consists predominantly of the dark neutrons~\cite{Ibe:2018juk}.
If the masses of the dark quarks are dominated by $v_D$, the dark neutron and the dark proton significantly mix in the mass basis.
In this case, a constraint from the DM direct detection experiment is stringent irrespectively of the DM constituent.

\subsection{Remarks}\label{sec:remarks}
Several comments are in order.
As we mentioned in \cref{sec:portal}, we assume that the dark electron gets a mass comparable to the dark GUT scale in order to ensure the longevity of ADM.
The decay of dark nucleons in Eq.\,(\ref{eq:nucdecay}) is kinematically prohibitted if the dark electron is heavier than the dark nucleons.%
\footnote{For example, we may extend the dark GUT model so that $\Xi'$ is complex scalar in the adjoint representation of $SU(4)_{\mathrm{DGUT}}$
with additional vector-like fermions $(X',\ovl{X}')$ in the $(\mathbf{4}, \ovl{\mathbf{4}})$ representations with a mass $M_X$.
Then, by assuming a softly broken chiral symmetry with a charge assignment $\Xi'(+1)$, $Q_{D}'(-1)$, $\ovl Q_{\ovl D}'(-1)$ and $Q'_{U}(+1)$,
the couplings of the dark quarks to $\Xi'$'s are restricted to  $y Q_{D}'\Xi'\ovl{X} + y X' \Xi' \ovl{Q}'_{\ovl D}  + M_X X' \ovl{X}' + \hc$
in the chiral symmetric limit.
Then, the masses of $D'$'s and $E'$'s are given by, $y^2v_{15}^2/M_X$ and $9y^2v_{15}^2/M_X$ for $M_X \gg v_{15}$, respectively,
which allows us to have about an order of magnitude larger dark electron mass than the dark quark masses.
By arranging the $y^2v_{15}^2/M_X$ in a sub-GeV to a few GeV range, the dark electron mass can be heavier than the dark nucleon masses,
with which the dark nucleon decay is prohibited.
}
In such a case, the dark electron can be much lighter than the right-handed neutrinos, and then the right-handed neutrinos can decay into the dark electron and the dark Higgs boson.
Such decays of the right-handed neutrinos can generate the $B - L$ asymmetry in addition to their decays into the visible lepton and the SM Higgs.
A new portal operator also arises below the energy scale of $M_R$,
\eqs{
\Lcal_{\text{new portal}} = \frac{y_N Y_N}{M_R} (\phi_D E') (LH) + \hc
\label{eq:new_portal}
}
This cosmology is intensively studied in Ref.~\cite{Falkowski:2011xh}.
This operator causes the dark electron decay into the SM neutrino with emitting the dark photon.

It might be tempting to consider the dark GUT model based on $SU(5)_{\mathrm{DGUT}}$,
where the dark quarks in  \cref{tab:Charge_dark} are unified into the $\mathbf{10}$ and the $\ovl{\mathbf 5}$ representations of $SU(5)_{\mathrm{DGUT}}$, instead of $SU(4)_{\mathrm{DGUT}}$.
However, it is difficult to make such a mirror model in the dark sector cosmologically viable.
The dark neutrinos in $SU(5)_\mathrm{DGUT}$ are massless up to lepton number violating operators as in the SM, and then they affect the effective number of neutrino species
$N_{\mathrm{eff}}$ and can behave like a hot component of DM as the SM neutrinos.
Furthermore, the dark nucleon can decay into the dark neutrino and the dark pion within the age of the Universe via dark nucleon decay operators if the $SU(5)_{\mathrm{DGUT}}$ GUT scale is close to $10^{10}\,\mathrm{GeV}$.
The latter problem cannot be avoided since the dark neutrino cannot be made heavier than the dark nucleons unlike the dark electron.

\section{Supersymmetric Realization \label{sec:susy}}
Intermediate-scale SUSY is theoretically and phenomenologically well-motivated UV physics (see Refs.~\cite{Nilles:1983ge, Haber:1984rc, Wess:1992cp, Martin:1997ns} for reviews).
It would be natural to consider a SUSY extension of our $SU(5)_{\mathrm{GUT}} \times SU(4)_{\mathrm {DGUT}}$ model.
Indeed, gauge couplings in the visible sector are precisely unified into one at the GUT scale when SUSY is assumed.
The non-renormalization theorem ensures our choices of the model parameters, such as fine-tuning for the GUT-scale splittings between $\phi_C$ and $\phi_D$ and between $D'$ and $E'$, against quantum corrections.
Scalar interactions are also restricted due to SUSY, and therefore some of Higgs multiplets are naturally light up to the little hierarchy between the SUSY breaking scale and the electroweak or the dark QED breaking scales.
Therefore, in this section, we consider a minimal SUSY realization of the composite ADM model.

We assume the minimal supersymmetric standard model (MSSM) in the visible sector, and a minimal SUSY extension of the dark sector.
$Q_U'\,, Q_D'\,,$ and $Q'_{\ovl D}$ are chiral superfields denoted by the same symbols as their fermionic components.
It should be noted that we introduce more than one generations of vector-like $Q_D'$ and $Q_{\ovl D}'$ so that the supersymmetric neutron chiral multiplets, i.e., $U'^a D'^b D'^c$
and $\ovl U'^a \ovl D'^b \ovl D'^c$, are available.
We take two generations in the following, although we suppress generation indices for the sake of notational simplicity.
We introduce two fundamental dark Higgs superfields, $H'$ and $\ovl H'$, and a dark adjoint superfield $\Xi'$.
One can refer to \cref{tab:Q5Chargeassign} for the charge assignment again.
The superpotential in the dark sector is given by
\eqs{
W & = \ovl H' (\mu + \lam \Xi') H' + W_{\Xi'} \\
& ~~ + Y_D \ep^{\alpha\beta\gamma\del} H'_{\alpha} Q'_{U[\beta\gamma]} Q'_{D \del}
+ Y_{\ovl D} \ovl H'^\alpha Q'_{U[\alpha \beta]} \ovl Q'_{\ovl D}{}^\beta
+ Y_N H'_{\alpha} \ovl Q'_{\ovl D}{}^\alpha \ovl N\,.
}
Here, $W_{\Xi'}$ denotes the superpotential including only $\Xi'$, and we assume that the superpotential is invariant under the $SU(5)_{\mathrm {GUT}} \times SU(4)_{\mathrm {DGUT}}$ and global $U(1)_5$ symmetries.
The $Q_U'\,, Q_D'\,,$ and $Q'_{\ovl D}$ have mass terms although they are not shown here.
As in the previous section, the masses of $\phi_C$ and $\phi_D$ should be split. %
The mass splitting between $\phi_C$ and $\phi_D$ is realized when fine-tuning of $\mu = 3 \lam v_{15}$ is assumed.%
\footnote{While we simply assume fine-tuned parameters in this section, we can naturally solve the mass splitting by introducing a non-minimal Higgs representation in SUSY models.
For instance, $\ovl{\mathbf{20}}$ representation in $SU(4)$ does not have $SU(3)_D$ singlet as a component, $\ovl{\mathbf{20}} \to \ovl{\mathbf{3}}_{1/3} + \mathbf{3}_{5/3} + \mathbf{6}_{1/3} + \mathbf{8}_{1}$.
Therefore, a product $\ovl{\mathbf{20}} \times \mathbf{4} = \mathbf{15} + \mathbf{20}^\prime + \mathbf{45}$ indicates that superpotential $W = H' \ovl\eta' \Xi' + \ovl H' \eta' \Xi'$, with $20$-dimensional chiral multiplets $\eta'(\mathbf{20})$ and $\ovl \eta'(\ovl{\mathbf{20}})$, gives a mass only for $\phi_C$ after the $SU(4)_{\mathrm {DGUT}}$ breaking.
This mechanism is similar as the missing-partner mechanism in $SU(5)_{\mathrm{GUT}}$ models~\cite{Grinstein:1982um, Masiero:1982fe}.
}

\subsection{Tiny Kinetic Mixing and $B-L$ Portal Operator}
The kinetic mixing in the SUSY model arises from
\eqs{
\Lcal & = \int d^2 \theta ~ \frac{1}{\MPl^2}\tr(\mathcal{W}_G \Si)  \Tr(\mathcal{W}_D \Xi') + \hc
\sim \int d^2 \theta ~ \frac{\ep}{4 \cos\theta_W} \mathcal{W} \mathcal{W}^\prime + \hc \, ,
}
where $\mathcal{W}$ and $\mathcal{W}^\prime$ are field strength chiral superfields of $U(1)_Y$ and $U(1)_D$, respectively.
The mixing parameter $\ep$ is defined in \cref{eq:epsilon}.

Below the energy scales of the masses of the dark colored Higgs triplet and the right-handed neutrinos, the following effective superpotential arises:
\eqs{
W_{\mathrm{eff.}} = - \frac{Y_N y_N Y_{\ovl D}}{M_C M_R} \ep_{abc} \ovl U'^a \ovl D'^b \ovl D'^c (L H_u)
\,,
\label{eq:portal_susy}
}
where $H_u$ is one of the MSSM Higgs doublets.
Again it should be noted that we introduce two generations of $D'$ and $\ovl D'$ so that the portal interaction does not vanish.
As in the case with a non-SUSY model, we simply assume that the dark electron chiral multiplet gets a mass of the order of the dark GUT scale in order to stabilize the dark neutron, and thus we omit the term like \cref{eq:new_portal}.

Due to superpartners of dark fermions, the portal interaction arises at dimension six rather than dimension seven.
This relaxes the phenomenological constraints \cref{eq:pheno_constraint,eq:Y_constraint} as
\eqs{
M_R < M_C \lesssim 10^2 ~ Y_N Y_{D , \ovl D} M_R \,, ~~~\text{and thus}~~~ Y_N Y_{D , \ovl D} \gtrsim 0.01 \,,
}
where we assume that $\tan \beta$, which is the ratio of the VEVs of the two Higgs doublets in the MSSM, is of the order of unity.

\subsection{Lightest Supersymmetric Particles in Two Sectros}

In SUSY extensions, the lightest supersymmetric particles (LSPs) would also be stable due to the R-parity (i.e., a discrete subgroup of $U(1)_5$).
In particular, we have two species of the LSPs both in the MSSM and in the dark sector.
They could lead to the overclosure of the Universe.
Even if their fractions to the total DM abundance are subdominant, the late-time (over one second) decay of the heavier LSP into the lighter one can cause cosmological problems.
The ratios are severely constrained by the big-bang nucleosynthesis and the spectral distortion of the cosmic microwave background, especially when their decay products are electromagnetic charged (see, e.g., Refs.~\cite{Poulin:2016anj, Kawasaki:2017bqm}).

In our setup, such harmful late-time decays could take place since the MSSM and the dark sectors feebly interact with each other below the energy scale of the order of $M_R$.
Indeed, there is no renormalizable interaction term between the visible and the dark sectors below it, if one turns off the kinetic mixing between the $U(1)_Y$ and the $U(1)_D$ vector multiplets.
For instance, one may think that the LSPs would be harmless if we make the dark squark the LSP in the dark sector and reduce its relic abundance through its efficient annihilation.
However, its lifetime induced by the operator \cref{eq:portal_susy} is too long:
\eqs{
\tau(\widetilde Q' \to Q'Q' \widetilde L H_u)
\sim \frac{2048 \pi^5 \Lam^4}{m_{\widetilde Q'}^5}
\sim 4 \times 10^6 ~ \mathrm{sec} \left( \frac{\Lam}{10^{10}\,\mathrm{GeV}} \right)^4 \left( \frac{m_{\widetilde Q'}}{1\,\mathrm{TeV}} \right)^{-5} \,,
}
where $m_{\widetilde Q}$ is the dark squark mass, and $\Lam^2 = M_C M_R / Y_N y_N Y_{\ovl D}$.

Thus the kinetic mixing between the $U(1)_Y$ and the $U(1)_D$ vector multiplets plays two important roles in making our cosmological scenario viable.
Its bosonic part, the kinetic mixing between the dark and the visible photons, transfers the dark sector entropy into the visible sector.
Its fermionic part, the kinetic mixing between the bino and the dark photino, which are fermionic partners of the $U(1)_Y$ gauge boson and the dark photon, respectively, helps the heavier LSP decay into the lighter one with a sufficiently short lifetime.

If the pure bino and the dark photino are the LSPs in the MSSM and the dark sectors, respectively, their relic abundance tends to be overabundant.
We, therefore, consider that the MSSM higgsinos are the LSP in the MSSM sector%
\footnote{The LSP in the MSSM sector can be the neutral wino instead.
}
while the dark higgsino, which is the fermionic partner of $U(1)_D$ breaking Higgs $\phi_D$, is the LSP in the dark sector.
More specifically, we take a split spectrum of sparticles \cite{ArkaniHamed:2004fb,Giudice:2004tc,Wells:2004di,Ibe:2006de} for simplicity: the gauginos and the higgsino have masses of $\Ocal(1)\,\mathrm{TeV}$ while all the scalar particles other than the SM Higgs and the $U(1)_D$ breaking Higgs are much heavier than $\Ocal(10^2)\,\mathrm{TeV}$.
In this case, the dark higgsino decay into the MSSM higgsino via the bino-dark photino kinetic mixing:
\eqs{
\tau(\widetilde \phi_D \to \phi_D H \widetilde H)
& \sim \frac{8 \pi}{\ep^2 \alpha_Y \alpha^\prime m_{\widetilde \phi_D}} \\
& \sim 2 \times 10^{-5} ~ \mathrm{sec} \left( \frac{10^{-9}}{\ep} \right)^2 \left( \frac{8 \times 10^{-2}}{\alpha^\prime} \right) \left( \frac{1~\mathrm{TeV}}{m_{\widetilde \phi_D}} \right) \, ,
}
where $m_{\widetilde \phi_D}$ is the mass of the dark higgsinos, and $\alpha_Y \simeq 0.01$ is the fine-structure constant of $U(1)_Y$.
We obtain the low-energy value of $U(1)_D$ fine structure constant, i.e., $\alpha^\prime \sim 8 \times 10^{-2}$, by setting $b^\prime = 2$ and $b_s^\prime = - 7$ instead in \cref{eq:U1Dcoupling} in the SUSY $SU(3)_D \times U(1)_D$ dynamics.
As a result, we find that the lifetime of the dark LSP decay through the kinetic mixing of the bino and the dark photino is much shorter than one second even though the mixing parameter is tiny.

Lastly, we comment on the dark LSP decay in the case with a light dark electron.
The dark higgsino $\widetilde \phi_D$ can decay into the dark electron, the slepton, and the MSSM Higgs doublet through the SUSY version of \cref{eq:new_portal}.
This decay process occurs faster than the decay through the kinetic mixing if the sleptons have the mass of the order of TeV.

\section{Conclusion \label{sec:conclusion}}
Composite ADM is an intriguing framework that naturally explains why the observed DM mass density is close to the baryon mass density.
On the other hand, the mechanism requires both high-energy and low-energy portals between the visible and the dark sectors.
The former transfers the asymmetry generated in one sector to the other, while the latter releases the resultant entropy of the dark sector into the visible sector.
We have constructed UV completions of a composite ADM model to clarify the origin of the two portals.

Our model is based on an $SU(5)_{\mathrm{GUT}} \times SU(4)_{\mathrm {DGUT}}$ gauge dynamics.
We have chosen the minimal candidate, i.e., $SU(4)_{\mathrm {DGUT}}$ gauge dynamics, for GUT in the dark sector.
$SU(4)_{\mathrm {DGUT}}$ is broken into $SU(3)_D \times U(1)_D$ at an intermediate scale.
$SU(3)_D$ provides dark hadrons, including dark nucleons as the ADM candidate.
Meanwhile the decay of the $U(1)_D$ dark photon releases the entropy of the dark sector into the visible sector.
The minimal dark quark contents are incorporated into vector-like representations of $SU(4)_{\mathrm {DGUT}}$, i.e., $\mathbf{4}+\ovl{\mathbf{4}}+\mathbf{6}$.

We have introduced a fundamental scalar field, whose $SU(3)_D$ singlet component develops a VEV to give a mass to the dark photon.
The high-energy portal interaction is mediated by the $SU(3)_D$ triplet component of the fundamental scalar and the heavy right-handed neutrinos.
Thanks to the global $U(1)_5$ symmetry, which is a GUT compatible extension of the $B-L$ symmetry, we can prohibit dangerous operators that carry different $B-L$ charges.

The low-energy portal, i.e., the kinetic mixing between the dark photon and the visible photon, is forbidden at the renormalizable level since both the $U(1)$ photons are parts of larger non-Abelian gauge bosons.
Above the GUT scale, we have na\"ively expected that all non-renormalizable operators are suppressed by the Planck mass.
We have obtained the preferable kinetic mixing $\ep \sim 10^{-9}$ when the dark GUT scale is set to be about $10^{10}\,\mathrm{GeV}$.
The mass of the $SU(3)_D$ triplet component of the fundamental scalar is of the order of this dark GUT scale and is compatible with thermal leptogenesis as a production mechanism of the $B-L$ asymmetry.

We have also considered the SUSY extension of the UV model since SUSY plays an important role in the gauge coupling unification in the visible sector and the stability of the GUT-scale mass splittings against quantum corrections.
However, the LSPs in both the sectors are long-lived and therefore could cause cosmological problems through their relic abundance and the late-time decay of the heavier LSP to the lighter LSP.
We have found that a SUSY version of the above kinetic mixing plays another important role here.
It leads to the kinetic mixing between the bino and the dark photino, through which the heavier LSP can decay into the lighter one with the lifetime much shorter than one second.
Especially when we consider a split spectrum of sparticles, i.e., light gauginos, light higgsinos, and heavy scalars, the LSPs are cosmologically harmless.

\subsection*{Acknowledgement}
This work is supported in part by Grants-in-Aid for Scientific Research from the Ministry of Education, Culture, Sports, Science, and Technology (MEXT) KAKENHI, Japan, No. 15H05889, No. 16H03991, No. 17H02878, and No. 18H05542 (M.I.) and by the World Premier International Research Center Initiative (WPI), MEXT, Japan.
The work of A. K. and T. K. is supported by IBS under the project code, IBS-R018-D1.
A. K. would like to acknowledge the European Centre for Theoretical Studies in Nuclear Physics and Related Areas (ECT*) for its hospitality during the completion of this work.

\appendix

\bibliographystyle{utphys}
\bibliography{draft_replacement}
\end{document}